# How to Beat the Adaptive Multi-Armed Bandit

Varsha Dani[*]   Thomas P. Hayes[†]

February 13, 2006

**Abstract**

The multi-armed bandit is a concise model for the problem of iterated decision-making under uncertainty. In each round, a gambler must pull one of $K$ arms of a slot machine, without any foreknowledge of their payouts, except that they are uniformly bounded. A standard objective is to minimize the gambler's *regret*, defined as the gambler's total payout minus the largest payout which would have been achieved by any fixed arm, *in hindsight*. Note that the gambler is only told the payout for the arm actually chosen, not for the unchosen arms.

Almost all previous work on this problem assumed the payouts to be *non-adaptive*, in the sense that the distribution of the payout of arm $j$ in round $i$ is completely independent of the choices made by the gambler on rounds $1, \ldots, i-1$. In the more general model of *adaptive* payouts, the payouts in round $i$ may depend *arbitrarily* on the history of past choices made by the algorithm.

We present a new algorithm for this problem, and prove nearly optimal guarantees for the regret against both non-adaptive and adaptive adversaries. After $T$ rounds, our algorithm has regret $O(\sqrt{T})$ with high probability (the tail probability decays exponentially). This dependence on $T$ is best possible, and matches that of the full-information version of the problem, in which the gambler is told the payouts for all $K$ arms after each round.

Previously, even for non-adaptive payouts, the best high-probability bounds known were $O(T^{2/3})$, due to Auer, Cesa-Bianchi, Freund and Schapire [1]. For non-adaptive payouts, they also proved an $O(\sqrt{T})$ bound on expected regret. We describe an adaptive payout scheme for which the expected regret of their algorithm is $\Omega(T^{2/3})$.

## 1 Introduction

In problems of "online decision making," a sequence of choices must be made without knowledge of the future. Typically, each decision results in a certain "cost" or "reward," which is immediately revealed to the algorithm for use in later decision-making.

In this setting, a common goal is to minimize the "regret" of the algorithm, defined as the algorithm's net cost minus the least cost, which would, *in hindsight*, have been incurred by a different decision sequence. This is clearly a hopeless task unless the decision sequences the algorithm will compare itself against are restricted somehow. Often this is achieved by only considering strategies that make a single decision and use it every round.

---

[*]Department of Computer Science, University of Chicago, Chicago IL 60637, varsha@cs.uchicago.edu.

[†]Computer Science Division, University of California, Berkeley CA 94720-1776, hayest@cs.berkeley.edu. Supported by an NSF Postdoctoral Fellowship and by NSF grant CCR-0121555.

Imagine a slot machine with $K$ arms. A gambler plays the slot machine repeatedly, each time by pulling one of the arms and paying its cost for that round. At the end of the day, the gambler wants his total cost to be not much more than that of the best single arm *in hindsight*.

We will focus on two different models of feedback to the gambler. In the *full information* setting, the costs of all $K$ arms are revealed to the gambler after each round. In the *bandit* setting, only the cost of the chosen arm is revealed, making life more difficult for the gambler.

## 1.1 Adaptive and non-adaptive cost allocation

Much of the early work in this area (see, e.g., [12, 6]) was based on the assumption that each machine has a time-invariant distribution, from which its cost is sampled independently in each round. In that case, the gambler's problem can be viewed as "learning" these distributions, while avoiding the unfavorable machines.

Subsequently, many papers have removed the assumption that the cost distribution is fixed, instead allowing it to vary arbitrarily over time. However, most of these results still assume that the way the costs vary is *completely unaffected* by the choices made by the gambler. We will refer to this model as *non-adaptive* cost allocation. Note that non-adaptive cost allocations are not required to be deterministic, and hence include time-invariant cost allocation as a special case.

In both of the above settings, essentially everything is known about the expected regret. The best possible expected regret in the full information version is known to be $\Theta(\sqrt{T \log K})$ [4, 5, 8] and in the bandit version, $O(\sqrt{TK \log K})$ [1, 2] and $\Omega(\sqrt{TK})$. The lower bounds hold even for time-invariant cost distributions, and the upper bounds hold for any non-adaptive cost allocation.

It is also natural to consider, for a parameter $\varepsilon > 0$, what are the best regret bounds which can be guaranteed with probability at least $1 - \varepsilon$. For this stronger type of guarantee, the answer does not change much for the full information version; the right answer is $\Theta(\sqrt{T(\log K + \log 1/\varepsilon)})$. However, for the bandit version of the problem, the best previously known high probability bounds were of the form $O(T^{2/3}(K \log K)^{1/3})$, even for $\varepsilon = 1/3$. We will improve these to $O(\sqrt{TK \log K} \log 1/\varepsilon)$.

In the more general setting of *adaptive* cost allocation, the costs are additionally allowed to depend on the decisions made by the gambler in previous rounds. The only independence requirement is that the gambler is allowed to randomize his current decision independently of the costs being set for the current round.

The best known lower bounds in the adaptive costs framework are the same (at least up to constant factors) as in the non-adaptive framework. In the full-information setting, adaptive payout cannot force any more expected regret than non-adaptive payouts can (see [3, Theorem 3.1]). In that setting, the optimal expected regret is $\Theta(\sqrt{T \log K})$.

In the bandit version however, adaptive payouts are strictly more powerful than non-adaptive. This can be seen by a brute force calculation of the optimal expected regret for the tiny example $K = 2, T = 2$. Here the worst expected regret that can be forced by adaptive costs is $2/3$, whereas non-adaptive costs can only force expected regret $1/2$. The present paper addresses the question, "How much more regret can an adaptive cost sequence force?" As we shall demonstrate, the answer is, "At most a constant factor."

In the bandit setting, the best previously known upper bound was $O(T^{2/3}(K \log K)^{1/3})$, due to Auer, Cesa-Bianchi, Freund and Schapire [1]. Our main contribution is a new algorithm for the bandit model which improves this upper bound to $O(\sqrt{TK \log K})$. This algorithm, which we call "Accounts," is stated in Section 2.

**Theorem 1.1.** *Let $R$ denote the regret of the "Accounts" algorithm for the $K$-armed bandit, on any adaptively chosen cost sequence of length $T$. Then, for every $\alpha > 1$,*

$$\mathbf{Pr}\left(R \geq (\alpha + 7)\sqrt{TK \ln K}\right) \leq 1000K\sqrt{\alpha}\exp\left(-\frac{\sqrt{\alpha}\log K}{8}\right).$$

*It follows that*

$$\mathbf{E}(R) = O(\sqrt{TK \ln K}).$$

None of the constants in Theorem 1.1 is tight. In particular, the $\sqrt{\alpha}$ controlling the rate of exponential decay can be replaced by $\alpha^{1-\varepsilon}$, for any $\varepsilon > 0$, at some expense to the other parameters. Our emphasis in this paper is on the main new theoretical ideas.

It is easy to see that every algorithm for the $K$-armed bandit problem has expected regret $\Omega(\sqrt{TK})$ for a time-invariant cost allocation scheme in which one of the machines pays 1 with probability $1/2 + \sqrt{K/T}$ and 0 otherwise, and the other $K - 1$ machines all pay 1 and 0 each with probability $1/2$. Roughly speaking, the random noise from the coin flips is enough to hide the identity of the best machine for about $T/K$ observations; meanwhile, because we are in the bandit model, there are only enough rounds to observe each machine about $T/K$ times. This shows that our upper bound on expected regret is within a $O(\sqrt{\log K})$ factor of being optimal.

We also show that the regret bound on the **Exp3** algorithm of Auer *et al.* [1] cannot be substantially improved for adaptive costs, resolving an open question from that paper.

**Theorem 1.2.** *For any $K \geq 2$ and parameters $\eta = \eta(T)$, $\gamma = \gamma(T)$, there exists an adaptive cost schedule such that*

$$\mathbf{E}(R) = \Omega(T^{2/3}),$$

*where $R$ denotes the regret for the **Exp3** algorithm with "exploration probability" $\gamma$ and "sensitivity" $\eta$.*

We prove Theorem 1.2 in Section 8.

**Remark 1.3.** Auer *et al.* [2] incorrectly asserted that **Exp3** has expected regret $O(\sqrt{T})$ for adaptive costs. What is true is that, if

$$R_j := \sum_{t=1}^{T} x^t \cdot c^t - \mathbf{e}_j \cdot c^t.$$

denotes the "regret against arm $j$," then

$$\max_j \mathbf{E}(R_j) = O(\sqrt{T})$$

holds for adaptive cost schedules. This expression equals the expected regret in the case of non-adaptive costs (where the index $j$ achieving the maximum is deterministic). However, in general the inequality

$$\mathbf{E}(R) = \mathbf{E}\left(\max_j R_j\right) \geq \max_j \mathbf{E}(R_j)$$

may be very far from equality, as observed by, among others, McMahan and Blum [11, Appendix B].

## 1.2 The "Accounts" algorithm

Our algorithm makes use of a "multiplicative update rule," selecting each arm randomly with probabilities that evolve based on their past performance. This principle is well-established in the literature; indeed, our algorithm can be seen as a direct descendant of the **Exp3** algorithm of Auer *et al.* [1]. That algorithm can in turn be viewed as the "natural" modification of the **Hedge** algorithm of Freund and Schapire [4] to the bandit setting. The Hedge algorithm is itself closely related to several earlier algorithms (see, e. g., Littlestone and Warmuth [9] and Vovk [13]). Indeed, as recently observed by Kalai [7], the Hedge algorithm can be seen as a special case of the "Follow the Perturbed Leader" algorithm (where the components of the perturbation are logarithms of exponentially distributed random variables).

The first step in modifying a full-information algorithm for use in the bandit setting is to devise a way to accurately guess the missing information. By "missing information" we do not mean the entire sequence of cost vectors so far (which would be impossible to estimate well), but rather the *sum* of cost vectors so far, since this is all the full-information algorithm needs. This is achieved by constructing a random variable whose expectation is the true cost vector, and whose value is computable by the algorithm from the cost of the one randomly chosen arm. Chernoff-Hoeffding bounds ensure that the sum of these random variables almost surely converges to the sum of the true cost vectors.

Unfortunately, when the probability of choosing a particular arm becomes small, the variance in the estimate of that arm's cost becomes large, which hurts the convergence rate for the sum. On the other hand, in order to approach the performance of the best arm, the algorithm needs to decrease its probabilities for choosing arms of higher cost. The **Exp3** algorithm of Auer *et al.* [1] balances these considerations by establishing a mixed "minimum exploration rate" of $\gamma/K$, but using the multiplicative weights rule to allocate the remaining $1 - \gamma$ probability.

Our algorithm instead has, for each arm, a sliding minimum exploration rate, which starts at approximately $1/K$, and may decrease to $O(1/\sqrt{T})$, depending on the costs encountered. This minimum exploration rate is enforced by means of an account $A_j$, for each arm $j$. Roughly speaking, the account fills up account $A_j$ with "negative regret" for its performance relative to arm $j$. The minimum exploration rate $g(A_j)$ is defined in such a way that the cumulative regret caused by the variance due to exploring at rate $g(A_j)$ over the remaining rounds will almost surely be less than the negative regret already stored in $A_j$.

Put another way, when the exploration probabilities are all above the minimum rates, the algorithm updates these probabilities using the usual multiplicative updates rule. If this rule results in a probability dropping too low, then instead of decreasing the exploration probability further, the algorithm instead adds the estimated cost vector into the account for that arm, and keeps the exploration probability the same. This has the result of initially elevating the exploration probabilities, but eventually letting them drop, for arms that perform consistently badly.

## 2 The model

We model the problem as a two-player zero-sum game between a gambler and a (rigged) casino. The number of rounds $T$ is fixed in advance, and known to both players. In each round $i$, the gambler chooses one arm $M^i \in \{1, \ldots, K\}$, while the casino simultaneously chooses costs $c_1^i, \ldots, c_K^i$ in some fixed bounded interval, which for notational convenience we take to be $[0, 1]$. After these choices are made, the casino is informed of the gambler's choice $M^i$, and the gambler is informed of the cost $c^i_{M^i}$ for his chosen arm. At the end of the game, the gambler's *loss* is defined as $\sum_{i=1}^T c^i_{M^i}$. The gambler's *regret* is defined as the difference

$$R = \sum_{i=1}^T c^i_{M^i} - \min_j \sum_{i=1}^T c^i_j.$$

Since we are interested in minimizing regret, we may think of $R$ as the payout to the casino, and $-R$ as the payout to the gambler.

## 3 The Algorithm

Let $S \subset \mathbb{R}^K$ denote the simplex of probability distributions over $\{1, \ldots, K\}$. Our algorithm is defined in terms of two functions $f : \mathbb{R}^K \to S$ and $g : \mathbb{R}_{\geq 0} \to [0, 1]$. The boldface variables are vectors in $\mathbb{R}^K$.

---

**Algorithm 3.1:** ACCOUNTS($f, g$)

$\widehat{\mathbf{C}} := \mathbf{A} := \mathbf{0}$.
**for** $i := 1$ **to** $T$
  Set $\mathbf{p} = (p_1, \ldots, p_K) = f(\widehat{\mathbf{C}})$.
  Sample $M = M^i$ from $1, \ldots, K$ according to the distribution $\mathbf{p}$.
  Pull arm $M$. Observe and incur cost $c^i_M$.
  **if** $g(A_M) \leq p_M$
    **then** $\widehat{\mathbf{C}}_M := \widehat{\mathbf{C}}_M + \frac{c^i_M}{p_M}$
  **else** $A_M := A_M + \frac{c^i_M}{p_M}$

---

Henceforth, we will work with the following specific choice of $f$. Let $\eta = \sqrt{\ln K / TK}$. For $z = (z_1, \ldots, z_K) \in \mathbb{R}^K$, and $j \in \{1, \ldots, K\}$, let

$$f_j(z) = \frac{e^{-\eta z_j}}{\sum_{\ell=1}^K e^{-\eta z_\ell}}.$$

We define our barrier function $g$ by

$$g(x) = \max\left\{\eta, \frac{1}{K(1 + x/\theta)^{3/2}}\right\},$$

where $\theta = \sqrt{KT \ln K}$. As will be fairly easily seen from our proof, if one's goal is only to derive an $O(\sqrt{T})$ bound on expected regret, the exponent $3/2$ may be replaced by any other value strictly between 1 and 2.

The innovative part of our algorithm is the introduction of the "accounts" vector $\mathbf{A}$, as well as the "moving barriers" $g(A_j)$. Without this device, some of the exploration probabilities could be made too small, resulting in too-high variance for the behavior of the algorithm compared to the arms of small probability. The basic idea is that the barriers enforce a lower bound on exploration probabilities for a given arm until the corresponding account has accumulated enough "negative regret" compared to that arm to "pay in advance" for the higher variance that may result after the barrier is lowered. Fortunately for us, the barriers can be lowered quite quickly and still achieve this goal, so that all of the barrier-created exploration combined is highly unlikely to result in more than $O(\sqrt{TK})$ total cost to the algorithm.

## 3.1 Conventions

In our analysis, we will use a compact notation to refer to the values taken on by the variables of our algorithm during the sequence of trials. For $0 \leq i \leq T$, a superscript $i$ will indicate the value of the variable at the end of round $i$. Thus for instance, $p_j^i = f_j(\widehat{\mathbf{C}}^{i-1})$ for $1 \leq i \leq T$. We consider all variables to have value 0 at the end of round 0.

For $j \in \{1, \ldots, K\}$, $i \in \{0, \ldots, T\}$, let $R_j^i$ denote "regret with respect to arm $j$ at time $i$," defined as

$$R_j^i = \sum_{\ell=1}^{i} c_{M^\ell}^\ell - c_j^\ell.$$

Note that this gives us a new formula for the final regret, $R$, namely,

$$R = \max_{1 \leq j \leq K} R_j^T.$$

For $j \in \{1, \ldots, K\}$, let $\Phi_j$ denote the following function from $\mathbb{R}^K \to \mathbb{R}$.

$$\Phi_j(\mathbf{z}) := \frac{1}{\eta} \ln \frac{1}{f_j(\mathbf{z})} = \frac{1}{\eta} \ln \frac{\sum_{\ell=1}^{K} e^{-\eta z_\ell}}{e^{-\eta z_j}} = z_j + \frac{1}{\eta} \ln \left( \sum_{\ell=1}^{K} e^{-\eta z_\ell} \right).$$

This definition implies that, for each $j$, $\nabla \Phi_j = \mathbf{e}_j - f$, i.e., for each $i, j$,

$$\frac{\partial}{\partial z_i} \Phi_j(\mathbf{z}) = \begin{cases} 1 - f_i(\mathbf{z}) & \text{if } i = j \\ -f_i(\mathbf{z}) & \text{otherwise.} \end{cases}$$

The motivation for defining $\Phi_j$ is that it acts as a potential function, controlling the change in $R_j$. This potential function is used in standard proofs of regret bounds for the weighted majority algorithm and variants. Let $\Phi_j^i = \Phi_j(\widehat{\mathbf{C}}^i)$, where $\widehat{\mathbf{C}}^i$ is the estimate for the sum of the costs after round $i$. One can easily see that $\Phi_j^0 = \frac{1}{\eta} \ln K$ and $\Phi_j^T \geq 0$. Thus the decrease $\Phi_j^0 - \Phi_j^T$ in potential over the entire game is bounded above by $\frac{\ln K}{\eta}$.

We will use $\Delta$ to denote the difference operator. Specifically, for $j \in [K]$ and $i \in [T]$, we will denote

$$\Delta R_j^i := R_j^i - R_j^{i-1}$$
$$\Delta \Phi_j^i := \Phi_j^i - \Phi_j^{i-1}$$
$$\Delta A_j^i := A_j^i - A_j^{i-1}.$$

Within the context of our proof of Theorem 1.1, for a particular value of $a$, we will restrict our attention to a fixed adaptive adversary, whose strategy maximizes $\mathbf{Pr}(R > a)$. This adversary may be assumed to be deterministic, in the sense that each cost function $c^i$ is a deterministic function of the previous decisions $M^1, \ldots, M^{i-1}$ made by the algorithm. We will denote by $\mathcal{H}_i$ the "history of the game prior to round $i$," that is, the $\sigma$-algebra generated by $M^1, \ldots, M^{i-1}$.

We also introduce the notations

$$Y_j^i := \Delta R_j^i + \Delta \Phi_j^i + \Delta A_j^i - \mathbf{E}\left(\Delta R_j^i + \Delta \Phi_j^i + \Delta A_j^i \,\big|\, \mathcal{H}_i\right),$$

$$Y = Y_j := \sum_{i=1}^T Y_j^i.$$

Note that the definition of $Y_j^i$ depends on the particular adversary in question. Also note that $Y_j^i$ is a martingale difference sequence.

We will use $\mathbf{e}_1, \ldots, \mathbf{e}_K$ to denote the standard basis for $\mathbb{R}^K$.

## 4 Outline of the Proof

The proof of our theorem splits into two main parts. First, we prove a weakly exponential tail inequality for the random variable $Y_j - A_j^T$. Essentially, this says that the account value $A_j^T$ rarely underestimates by much the contribution of "stepwise variance" to $R_j$.

**Lemma 4.1.** *Let $1 \leq j \leq K$. Then, for every $\alpha \geq 1$,*

$$\mathbf{Pr}\left(Y_j - A_j^T > (\alpha + 1)\sqrt{TK \ln K}\right) \leq \left(\frac{16\sqrt{\alpha}}{\ln K} + \frac{128}{\ln^2 K}\right) \exp\left(-\frac{\sqrt{\alpha} \ln K}{8}\right)$$

Second, we will prove that the contribution of "stepwise expectations" to $R_j + A_j^T$ exhibits an even sharper cut-off at $O(\sqrt{TK \log K})$.

**Lemma 4.2.**

$$\mathbf{Pr}\left(\exists j \ \sum_{i=1}^T \mathbf{E}\left(\Delta R_j^i + \Delta \Phi_j^{\,i} + \Delta A_j^i \,\big|\, \mathcal{H}_i\right) > 6\sqrt{TK \ln K}\right) \leq \exp\left(\frac{-3\sqrt{TK \ln K}}{26}\right)$$

In order to prove these lemmas, we will need some results about martingales which we will describe in Section 5. We will then prove Lemmas 4.1 and 4.2 in Sections 6 and 7 respectively. In the remainder of this section we will show how Theorem 1.1 follows from the two lemmas.

*Proof of Theorem 1.1.* Since $R$ cannot exceed $T$, we may assume without loss of generality that $(\alpha + 7)\sqrt{TK \ln K} < T$; otherwise, there is nothing to prove. Fix an arm $j$. By the definition of $Y_j$, we have

$$Y_j = \sum_{i=1}^T Y_j^i = R_j^T - R_j^0 + \Phi_j^T - \Phi_j^0 + A_j^T - A_j^0 - \sum_{i=1}^T \mathbf{E}\left(\Delta R_j^i + \Delta \Phi_j^{\,i} + \Delta A_j^i \,\big|\, \mathcal{H}_i\right).$$

Since $R_j^0 = A_j^0 = 0$ and $\Phi_j^0 - \Phi_j^T \leq \Phi(0) = \frac{\ln K}{\eta} = \sqrt{TK \ln K}$, this implies

$$R_j^T \leq Y_j - A_j^T + \sum_{i=1}^T \mathbf{E}\left(\Delta R_j^i + \Delta \Phi_j^{\,i} + \Delta A_j^i \,\big|\, \mathcal{H}_i\right) + \sqrt{TK \ln K}$$

Suppose the high probability bound of Lemma 4.2 holds, and that, for every $1 \leq j \leq K$, the high probability bound of Lemma 4.1 holds for $Y_j - A_j^T$. Then, summing, we obtain the desired bound on the regret,
$$R \leq \max_j R_j^T \leq (\alpha + 7)\sqrt{TK \ln K}.$$
Summing the error probabilities completes the proof for the tail inequality,
$$\mathbf{Pr}\left(R \geq (\alpha + 7)\sqrt{TK \ln K}\right) \leq K\left(\frac{16\sqrt{\alpha}}{\ln K} + \frac{128}{\ln^2 K}\right) \exp\left(-\frac{\sqrt{\alpha} \ln K}{8}\right) + \exp\left(\frac{-3\sqrt{TK \ln K}}{26}\right).$$
Noting that the second term is dominated by the first term, and approximating crudely using the facts $\alpha \geq 1$ and $K \geq 2$, we can infer that
$$\mathbf{Pr}\left(R \geq (\alpha + 7)\sqrt{TK \ln K}\right) \leq \frac{200K\sqrt{\alpha}}{\ln K} \exp\left(-\frac{\sqrt{\alpha} \ln K}{8}\right).$$
To prove the upper bound on expectation, we note that, in general,
$$\mathbf{E}(R) \leq \mathbf{E}(\max\{R, 0\}) = \int_0^\infty \mathbf{Pr}(R \geq x) \mathrm{d}x.$$
Using the trivial bound $\mathbf{Pr}(R \geq x) \leq 1$ for small values of $x$, and our main tail inequality for larger values of $x$, we deduce, after several steps, the desired estimate. $\square$

## 5 Concentration Inequalities for Martingales

The well-known Hoeffding-Azuma inequality bounds the probability of large deviation for a martingale with uniformly bounded step sizes. In our analysis, we will need rather tight bounds on the deviation probabilities for a martingale whose step sizes are bounded by $1/p^i$, which is a random variable. Fortunately, the quality of the bound attained is allowed to depend on the step sizes actually encountered.

The following strong generalization of the Hoeffding-Azuma inequality, due to McDiarmid [10, Theorem 3.15], does essentially what we want. In order to state his bound, we need to first introduce some concepts related to the notion of conditional expectation.

Recall that, for random variables $A$ and $B$ over a finite probability space $\Omega$, the conditional expectation $\mathbf{E}(A \mid B)$ is the random variable which, on each atom $\{B = b\}$ of $B$, takes value $\mathbf{E}(A \mid B = b)$, which is the expectation of $A$ in the corresponding restricted probability space. Analogously, we define the *conditional variance*, $\mathbf{Var}(A \mid B)$ to be the random variable which, on each atom $\{B = b\}$ of $B$, takes value $\mathbf{Var}(A \mid B = b)$, defined as the variance of $A$ in the corresponding restricted probability space. Again analogously, we define the *conditional positive deviation*, $\sup(A \mid B)$ to be the random variable which, on each atom $\{B = b\}$ of $B$, takes the maximum value attained by $A$ on that subset of $\Omega$. More generally, this is the minimum among all $B$-measurable random variables that are always $\geq A$. As usual, if $B = (B_1, \ldots, B_m)$ is a tuple of random variables, we will also write $\mathbf{Var}(A \mid B)$ as $\mathbf{Var}(A \mid B_1, \ldots, B_m)$, and also $\sup(A \mid B)$ as $\sup(A \mid B_1, \ldots, B_m)$.

**Theorem 5.1 (McDiarmid).** *Suppose $X_1, \ldots, X_n$ is a martingale difference sequence, and $b$ is an uniform upper bound on the steps $X_i$. Let $V$ denote the sum of conditional variances,*

$$V = \sum_{i=1}^{n} \mathbf{Var}\left(X_i \mid X_1, \ldots, X_{i-1}\right).$$

*Then, for every $a, v \geq 0$,*

$$\mathbf{Pr}\left(\sum X_i \geq a \text{ and } V \leq v\right) \leq \exp\left(-\frac{a^2}{2v + 2ab/3}\right).$$

We will also need the following more general formulation, which is an easy consequence.

**Theorem 5.2.** *Suppose $X_1, \ldots, X_n$, $V$, are as in Theorem 5.1. Let $B$ denote the maximum "conditional positive deviation,"*

$$B = \max_i \sup(X_i \mid X_1, \ldots, X_{i-1})$$

*Then, for every $a, b, v \geq 0$,*

$$\mathbf{Pr}\left(\sum X_i \geq a \text{ and } V \leq v \text{ and } B \leq b\right) \leq \exp\left(-\frac{a^2}{2v + 2ab/3}\right).$$

*Proof.* Define $X_1^*, \ldots, X_T^*$ inductively by setting $X_i^* = X_i$ unless $\sup(X_i \mid X_1, \ldots, X_{i-1}) > b$; in that case, set $X_i^* = \cdots = X_n^* = 0$. It is easily verified that $X^*$ is also a martingale difference sequence, and that $b$ is an absolute upper bound on the step sizes for $X^*$. Moreover, $X^*$ behaves exactly like $X$ except when $B > b$. Applying this together with Theorem 5.1 implies

$$\mathbf{Pr}\left(\sum X_i \geq a \text{ and } V \leq v \text{ and } B \leq b\right) \leq \mathbf{Pr}\left(\sum X_i^* \geq a \text{ and } V^* \leq v\right)$$

$$\leq \exp\left(-\frac{a^2}{2v + 2ab/3}\right),$$

where $V^*$ denotes $\sum_i \mathbf{Var}\left(X_i^* \mid X_1^*, \ldots, X_{i-1}^*\right)$. □

## 6 Charging for variance: Proof of Lemma 4.1

In this section, we prove Lemma 4.1, our tail inequality for $Y - A_j^T$. The key ingredient will be Lemma 6.6, which is a tail inequality for the "rectangular" events $\{Y_j \geq \zeta \text{ and } A_j^T \leq \xi\}$. First, however, we need to prove several basic facts, which will be used to prove upper bounds on the conditional variance and conditional positive deviation of the steps $Y_j^i$, in terms of the final account value $A_j^T$.

We first show that the probability of choosing arm $j$ cannot change drastically from one round to the next.

**Proposition 6.1.** *Let $1 \leq i < T$ and $1 \leq j \leq K$. If arm $j$ is chosen in round $i$, then*

$$e^{-\eta/p_j^i} \leq \frac{p_j^{i+1}}{p_j^i} \leq 1. \tag{1}$$

*Otherwise, if $M = M^i \neq j$ is the arm chosen in round $i$, then*

$$1 \leq \frac{p_j^{i+1}}{p_j^i} \leq e^{\eta c_M^i}, \tag{2}$$

*Proof.* Suppose $\widehat{\mathbf{C}}^{i-1} = (z_1, \ldots, z_K)$. Since $f_j$ is decreasing in its $j$'th component, and increasing in every other component, the inequalities involving 1 follow. Next, observe that when arm $j$ is chosen,

$$\frac{p_j^{i+1}}{p_j^i} = \frac{e^{-\eta(z_j + c_j^i/p_j^i)}}{e^{-\eta z_j}} \frac{\sum_\ell e^{-\eta z_\ell}}{\sum_\ell e^{-\eta(z_\ell + \delta_{\ell,j} c_j^i/p_j^i)}}$$

$$\geq e^{-\eta/p_j^i}.$$

When any other arm $M$ is chosen,

$$\frac{p_j^{i+1}}{p_j^i} = \frac{\sum_\ell e^{-\eta z_\ell}}{\sum_\ell e^{-\eta(z_\ell + \delta_{\ell,M} c_M^i/p_M^i)}}$$

$$= \frac{1}{1 - p_M^i(1 - e^{-\eta c_M^i/p_M^i})}$$

$$\leq e^{\eta c_M^i},$$

where the last inequality holds for every $0 \leq p_M^i \leq 1$ and $\eta c_M^i \geq 0$, with equality at $\eta c_M^i = 0$, as can be seen by examining the first partial derivative in $x = \eta c_M^i$. □

Next we bound changes in regret, potential, and account value in terms of the arm probability $p_j^i$. .

**Proposition 6.2.** *Let $1 \leq i \leq T$ and $1 \leq j \leq K$. If arm $j$ is chosen in round $i$, then*

$$0 \leq \Delta R_j^i + \Delta \Phi_j^i + \Delta A_j^i \leq 1/p_j^i$$

*When any other arm $M$ is chosen,*

$$-1 \leq \Delta R_j^i + \Delta \Phi_j^i + \Delta A_j^i \leq 0.$$

*Proof.* More precisely, we show that if arm $j$ is chosen in round $i$, then $\Delta R_j^i = 0$ and both $\Delta \Phi_j^i$ and $\Delta A_j^i$ are between 0 and $1/p_j^i$, with at most one of them nonzero. When an arm $M \neq j$ is chosen in round $i$, we show that $\Delta A_j^i = 0$, $\Delta R_j^i = c_M^i - c_j^i$ and $-c_M^i \leq \Delta \Phi_j^i \leq 0$.

The claims for $\Delta R_j^i$ are immediate from the definition. The bounds on $\Delta A_j^i$ follow because $A_j^i$ is only incremented when arm $j$ is chosen (and moreover $g(A_j^{i-1}) > p_j^i$), and then $\Delta A_j^i = c_j^i/p_j^i \leq 1/p_j^i$. To see the bounds on $\Delta \Phi_j^i$, note that

$$\Delta \Phi_j^i = \frac{1}{\eta} \ln \left( \frac{p_j^i}{p_j^{i+1}} \right),$$

Combining this with Proposition 6.1 completes the proof. □

As stated in the Introduction, the purpose of the accounts is to prevent the algorithm from overreacting to observed costs and allowing the corresponding probabilities to decrease too quickly. The next proposition states that, indeed, for each $j$, $g(A_j)$ acts as an approximate lower bound on the probability of choosing arm $j$. In particular, the exploration probabilities never drop below $\Omega(1/\sqrt{T})$. In the **Exp3** algorithm of Auer *et al.* [1, 2], the latter property was enforced by using a modified version of $f$.

**Proposition 6.3.** *For $1 \leq i \leq T$ and $1 \leq j \leq K$,*

$$e^{\eta/g(A_j^{i-1})} p_j^i \geq g(A_j^{i-1}) \geq g(A_j^T).$$

*Proof.* Since $g$ is decreasing and $A_j^i$ is increasing in $i$, we have $g(A_j^{i-1}) \geq g(A_j^T)$. We prove by induction on $i$ that $p_j^i \geq g(A_j^{i-1})e^{-\eta/g(A_j^{i-1})}$. The base case $i = 1$ is clear, since $p_j^1 = f_j(\mathbf{0}) = 1/K \geq g(0)$.

For the inductive step, we consider two cases. If $p_j^{i+1} \geq p_j^i$, then by inductive hypothesis we have

$$p_j^{i+1} \geq p_j^i \geq g(A_j^{i-1})e^{-\eta/g(A_j^{i-1})} \geq g(A_j^i)e^{-\eta/g(A_j^i)},$$

where the last inequality follows because $A_j^i \geq A_j^{i-1}$ and $g$ is non-increasing.

On the other hand, if $p_j^{i+1} < p_j^i$, then note that $j$ must be the arm chosen in round $i+1$ and moreover $p_j^i \geq g(A_j^{i-1})$. This also implies that $A_j^i = A_j^{i-1}$. It follows by Proposition 6.1 that

$$p_j^{i+1} \geq p_j^i e^{-\eta/p_j^i} \geq g(A_j^{i-1})e^{-\eta/g(A_j^{i-1})} = g(A_j^i)e^{-\eta/g(A_j^i)}. \qquad \square$$

Next, we show how to derive good upper bounds on the conditional variances and conditional positive deviations for our martingale steps $Y_j^i$.

**Proposition 6.4.** *For all $j, i$, $Y_j^i \leq e^{\eta/g(A_j^T)}/g(A_j^T)$.*

*Proof.* Let $Z_j^i$ denote $\Delta R_j^i + \Delta \Phi_j^i + \Delta A_j^i$. Note that by definition $Y_j^i = Z_j^i - \mathbf{E}\left(Z_j^i \,\middle|\, \mathcal{H}_i\right)$. Let $\mu = \sup(Z_j^i \mid \mathcal{H}_i)$ denote the conditional positive deviation of $Z_j^i$ given $\mathcal{H}_i$. By Proposition 6.2, $Z_j^i$ attains value $\mu$ when arm $j$ is chosen, which occurs with probability $p_j^i$, and moreover $\mu \leq 1/p_j^i$, and when any arm is chosen, $Z_j^i \geq -1$. It follows that

$$\mathbf{E}\left(Z \mid \mathcal{H}_i\right) \geq p_j^i \mu - (1 - p_j^i).$$

From this we infer

$$Y_j^i = Z^i - \mathbf{E}\left(Z \mid \mathcal{H}_i\right) \leq \mu - \mathbf{E}\left(Z \mid \mathcal{H}_i\right) \leq (1 - p_j^i)\mu + (1 - p_j^i) \leq 1/p_j^i - p_j^i < 1/p_j^i.$$

By Proposition 6.3, it follows that

$$Y_j^i \leq 1/p_j^i \leq e^{\eta/g(A_j^{i-1})}/g(A_j^{i-1}) \leq e^{\eta/g(A_j^T)}/g(A_j^T).$$

$\square$

**Proposition 6.5.** *For $1 \leq j \leq K$,*

$$\sum_{i=1}^{T} \mathbf{Var}\left(Y_j^i \mid \mathcal{H}_i\right) \leq T\left(1 + e^{\eta/g(A_j^T)}/g(A_j^T)\right).$$

*Proof.* Let $1 \leq i \leq T$, and condition on $\mathcal{H}_i$, which in particular determines $p = p_j^i$. By Proposition 6.2, we know that $Z = \Delta R_j^i + \Delta \Phi_j^i + \Delta A_j^i$ is between $0$ and $1/p$ with probability $p$, and otherwise between $-1$ and $0$. Since, for any random variable $X$, $\mathbf{Var}(X) \leq \mathbf{E}(X^2)$, this implies the conditional variance $\mathbf{Var}(Z \mid \mathcal{H}_i) \leq p(1/p)^2 + (1-p)(-1)^2 < 1/p + 1$. Since $Y_j^i$ equals $Z$ minus its conditional expectation, it has the same conditional variance. Thus $\mathbf{Var}\left(Y_j^i \mid \mathcal{H}_i\right) < 1 + 1/p$. By Proposition 6.3, this is at most $1 + e^{\eta/g(A_j^T)}/g(A_j^T)$. Summing over $T$ completes the proof. $\square$

Next we prove a tail inequality for the "rectangular" events $\{Y_j \geq \zeta \text{ and } A_j^T \leq \xi\}$. The key insight is that, via Propositions 6.4 and 6.5, we can apply McDiarmid's inequality to these events.

**Lemma 6.6.** *For all $\xi \geq 0$ and $1 \leq j \leq K$,*

$$\mathbf{Pr}\left(Y_j \geq \zeta \text{ and } A_j^T \leq \xi\right) \leq \exp\left(-\frac{\zeta^2 g(\xi)}{6T + 2\zeta}\right).$$

*Proof.* Fix $j$. By definition, $Y_j^1, \ldots, Y_j^T$ is a martingale difference sequence with respect to the filtration $\mathcal{H}_1, \ldots, \mathcal{H}_T$. Let $V$ denote the sum of conditional variances, $V = \sum_{i=1}^{T} \mathbf{Var}\left(Y_j^i \mid \mathcal{H}_i\right)$, and let $B$ denote the maximum conditional positive deviation,

$$B = \max_i \sup(Y_j^i \mid \mathcal{H}_i).$$

By Propositions 6.4 and 6.5, we know $B \leq e^{\eta/g(A_j^T)}/g(A_j^T)$ and $V \leq T(1 + e^{\eta/g(A_j^T)}/g(A_j^T))$. Consequently, for every $\zeta, \xi$, we know

$$\mathbf{Pr}\left(Y \geq \zeta \text{ and } A_j^T \leq \xi\right) \leq \mathbf{Pr}\left(Y \geq \zeta \text{ and } V \leq T(1 + e^{\eta/g(\xi)}/g(\xi)) \text{ and } B \leq e^{\eta/g(\xi)}/g(\xi)\right).$$

Applying Theorem 5.2 (McDiarmid's inequality), this implies

$$\mathbf{Pr}\left(Y \geq \zeta \text{ and } A_j^T \leq \xi\right) \leq \exp\left(-\frac{\zeta^2 g(\xi)}{2T(g(\xi) + e^{\eta/g(\xi)}) + 2e^{\eta/g(\xi)}\zeta/3}\right).$$

To simplify the denominator of the exponent on the right-hand side, note that it is a decreasing function of $g(\xi)$, and that by definition $g(\xi) \geq \eta$. Thus,

$$2T(g(\xi) + e^{\eta/g(\xi)}) + 2e^{\eta/g(\xi)}\zeta/3 \leq 2T(\eta + e) + 2e\zeta/3$$
$$\leq 6T + 2\zeta \qquad \text{since } \eta + e < 3.$$

Plugging this into the upper bound on $\mathbf{Pr}\left(Y \geq \zeta \text{ and } A_j^T \leq \xi\right)$ completes the proof. $\square$

Now we are ready to prove the main result of the section, Lemma 4.1. The proof combines Lemma 6.6 with a covering argument. We remark that, up to this point, the only properties of $g$ which have been used in an essential way are that it is decreasing and has minimum value $\eta$. Now we will begin to make use of the specific definition of $g$.

*Proof of Lemma 4.1.* Let $A_{\text{crit}} := \theta((\eta K)^{-2/3} - 1)$, which is the minimum $A$ such that $g(A) = \eta$. Let $A_{\max}$ denote the absolute maximum possible value of $A_j^T$. (Propositions 6.1 and 6.2 imply explicit bounds on $A_{\max}$, but we will not need these.) Let $\nu = (\alpha+1)\sqrt{TK \ln K}$.

Our approach is to cover the event $\{Y_j \geq A_j^T + \nu\}$ by rectangles of the form $\{Y_j \geq \zeta \text{ and } A_j^T \leq \xi\}$. More specifically, for $\ell \geq 1$, let $\zeta_\ell = (\alpha+\ell)\theta = \nu + (\ell-1)\theta$ and

$$\xi_\ell = \begin{cases} \ell\theta & \text{if } \ell\theta < A_{\text{crit}} \\ A_{\max} & \text{otherwise.} \end{cases}$$

Then the $\lceil A_{\text{crit}}/\theta \rceil$ rectangles $\{Y_j \geq \zeta_\ell \text{ and } A_j^T \leq \xi_\ell\}$, for $1 \leq \ell \leq \lceil A_{\text{crit}}/\theta \rceil$, cover the trapezoid $\{Y_j \geq A_j^T + \nu \text{ and } A_j^T \leq A_{\max}\}$, which equals the event $\{Y_j - A_j^T \geq \nu\}$.

*Claim.* For the above definition of $\zeta_\ell, \xi_\ell$,

$$\frac{\zeta_\ell^2 g(\xi_\ell)}{6T + 2\zeta_\ell} \geq \frac{(\alpha+\ell)^{1/2} \ln K}{8}.$$

*Proof of Claim.* We will show that the ratio

$$r := \frac{\zeta_\ell^2 g(\xi_\ell)}{(6T + 2\zeta_\ell)(\alpha+\ell)^{1/2}} \geq \frac{\theta^2}{K(6T + 2(\theta + A_{\text{crit}}))}.$$

First, observe that $r$ is an increasing function of $\alpha$, and hence we may assume $\alpha = 1$. In the case when $\ell\theta < A_{\text{crit}}$, this gives us

$$r = \frac{\theta^2}{K(6T + 2(1+\ell)\theta)} > \frac{\theta^2}{K(6T + 2(\theta + A_{\text{crit}}))}.$$

In the case when $\ell\theta \geq A_{\text{crit}}$, we have

$$r = \frac{(1+\ell)^{3/2}\theta^2 \eta}{6T + 2(1+\ell)\theta},$$

which differentiation shows is a decreasing function of $\ell$, and is hence minimized when $\ell = A_{\text{crit}}/\theta$. Substituting this, and using the fact that $g(A_{\text{crit}}) = \eta$, we have

$$r = \frac{(1 + A_{\text{crit}}/\theta)^{3/2}\theta^2 \eta}{6T + 2(\theta + A_{\text{crit}})} = \frac{\theta^2}{K(6T + 2(\theta + A_{\text{crit}}))}.$$

Next, observe that, since we are assuming $T \geq \sqrt{TK \ln K}$, it follows that

$$\theta + A_{\text{crit}} = T^{5/6}(K \ln K)^{1/6} \leq T.$$

Since $\theta^2 = TK \ln K$, this completes the proof of the Claim.

Applying Lemma 6.6 in conjunction with a union bound, we have

$$\mathbf{Pr}\left(Y_j - A_j^T \geq \nu\right) \leq \sum_{\ell=1}^{\lceil A_{\text{crit}}/\theta \rceil} \mathbf{Pr}\left(Y_j \geq \zeta_\ell \text{ and } A_j^T \leq \xi_\ell\right) \quad \text{by our covering argument}$$

$$\leq \sum_{\ell=1}^{\lceil A_{\text{crit}}/\theta \rceil} \exp\left(-\frac{\zeta_\ell^2 g(\xi_\ell)}{6T + 2\zeta_\ell}\right) \quad \text{by Lemma 6.6}$$

$$\leq \sum_{\ell=1}^{\lceil A_{\text{crit}}/\theta \rceil} \exp\left(-\frac{(\alpha+\ell)^{1/2} \ln K}{8}\right) \quad \text{by the Claim}$$

$$\leq \sum_{\ell=1}^{\infty} \exp\left(-\frac{(\alpha+\ell)^{1/2} \ln K}{8}\right)$$

$$\leq \int_{x=0}^{\infty} \exp\left(-\frac{(\alpha+x)^{1/2} \ln K}{8}\right) \mathrm{d}x$$

$$= \frac{128}{\ln^2 K} \int_{y=(\sqrt{\alpha}\ln K)/8}^{\infty} y \exp(-y) \mathrm{d}y$$

$$= -\frac{128}{\ln^2 K}(y+1)\exp(-y)\Big|_{y=(\sqrt{\alpha}\ln K)/8}^{\infty}$$

$$= \left(\frac{16\sqrt{\alpha}}{\ln K} + \frac{128}{\ln^2 K}\right)\exp\left(-\frac{\sqrt{\alpha}\ln K}{8}\right).$$

This completes the proof of Lemma 4.1. □

# 7 Bounding the sum of conditional expectations: Proof of Lemma 4.2

In this section, we prove Lemma 4.2, our tail inequality for the sum of conditional expectations $\mathbf{E}\left(\Delta R_j^i + \Delta \Phi_j{}^i + \Delta A_j^i \mid \mathcal{H}_i\right)$. We will first need the following bound on the accuracy of linear approximation to the potential function $\Phi_j$.

**Lemma 7.1.** *For $i \in [T], j \in [K]$, regardless of the history,*

$$\mathbf{E}\left(\Delta \Phi_j^i \mid \mathcal{H}_i\right) \leq \nabla \Phi_j(\widehat{\mathbf{C}}^{i-1}) \cdot \mathbf{E}\left(\Delta \widehat{\mathbf{C}}^i \mid \mathcal{H}_i\right) + \eta K.$$

*Proof of Lemma 7.1.* Let $i \in [T]$ be fixed. $\Delta \Phi_j^i = \Phi_j(\widehat{\mathbf{C}}^i) - \Phi_j(\widehat{\mathbf{C}}^{i-1})$. Once the history $\mathcal{H}_i$ has been fixed, $\widehat{\mathbf{C}}^{i-1}$ is determined. Let $\mathbf{z} = \widehat{\mathbf{C}}^{i-1}$, and let $\mathbf{z} + \Delta \mathbf{z} = \widehat{\mathbf{C}}^i$. Note that $\Delta \mathbf{z}$ is a random variable that takes value $\frac{c_\ell^i}{p_\ell^i}\mathbf{e}_\ell$ with probability $p_\ell^i$. We want to show that

$$\mathbf{E}\left(\Phi_j(\mathbf{z} + \Delta \mathbf{z}) - \Phi_j(\mathbf{z})\right) = \nabla \Phi_j(\mathbf{z}) \cdot \mathbf{E}(\Delta \mathbf{z}) + \eta K.$$

For any vector $\mathbf{v}$, let $\mathrm{D}_\mathbf{v}$ denote the (normalized) partial differential operator in the direction of $\mathbf{v}$. We abbreviate $\mathrm{D}_{\mathbf{e}_\ell}$ by $\mathrm{D}_\ell$. By Taylor's theorem, for some $\alpha \in [0, 1]$,

$$\Phi_j(\mathbf{z} + \Delta \mathbf{z}) - \Phi_j(\mathbf{z}) = \mathrm{D}_{\Delta \mathbf{z}} \Phi_j(\mathbf{z}) \|\Delta \mathbf{z}\| + \mathrm{D}_{\Delta \mathbf{z}}^2 \Phi_j(\mathbf{z} + \alpha \Delta \mathbf{z}) \|\Delta \mathbf{z}\|^2.$$

Taking expectations over $\Delta \mathbf{z}$, this becomes

$$\mathbf{E}\left(\Phi_j(\mathbf{z}+\Delta\mathbf{z})-\Phi_j(\mathbf{z})\right) = \nabla \Phi_j(\mathbf{z}) \cdot \mathbf{E}\left(\Delta \mathbf{z}\right) + \mathbf{E}\left(\mathrm{D}^2_{\Delta\mathbf{z}} \Phi_j(\mathbf{z}+\alpha\Delta\mathbf{z}) \left\|\Delta\mathbf{z}\right\|^2\right)$$

So we just need to show that the expectation of the second order term has the right bound.

Recall that $\nabla \Phi_j = \mathbf{e}_j - f$. Differentiating once more shows, for any vector $\mathbf{x}$,

$$\mathrm{D}^2_\ell \Phi_j(\mathbf{x}) = -\mathrm{D}_\ell f_\ell(\mathbf{x}) = \eta f_\ell(\mathbf{x})(1-f_\ell(\mathbf{x})) \leq \eta f_\ell(\mathbf{x}).$$

Applying this to the second-order term from Taylor's theorem yields

$$\mathbf{E}\left(\mathrm{D}^2_{\Delta\mathbf{z}}\Phi_j(\mathbf{z}+\alpha\Delta\mathbf{z})\left\|\Delta\mathbf{z}\right\|^2\right) = \sum_{\ell=1}^K p_\ell \, \mathrm{D}^2_\ell \Phi_j\left(\mathbf{z}+\alpha_\ell \frac{c_\ell}{p_\ell}\mathbf{e}_\ell\right)\left(\frac{c_\ell}{p_\ell}\right)^2$$

$$\leq \sum_{\ell=1}^K \eta \, f_\ell\left(\mathbf{z}+\frac{\alpha_\ell c_\ell}{p_\ell}\mathbf{e}_\ell\right) \frac{(c_\ell)^2}{p_\ell}$$

$$\leq \sum_{\ell=1}^K \eta \, \frac{f_\ell\left(\mathbf{z}+\alpha_\ell \frac{c_\ell}{p_\ell}\mathbf{e}_\ell\right)}{f_\ell(\mathbf{z})}$$

$$\leq \eta K,$$

where the last inequality follows because $f_\ell$ is decreasing in its $\ell^{\text{th}}$ component. $\square$

We now prove Lemma 4.2, our concentration bound on the sum of conditional expectations $\sum_i \mathbf{E}\left(\Delta R_j^i + \Delta \Phi_j^i + \Delta A_j^i \,\big|\, \mathcal{H}_i\right)$.

*Proof of Lemma 4.2.* For $1 \leq j \leq K, 1 \leq i \leq T$, let $b_j^i$ be the indicator variable for the event $\{g(A_j^{i-1}) \leq p_j^i\}$ (*i.e.,* the algorithm is not at the barrier for arm $j$ in round $i$). Note that $b_j^i$ is determined by $\mathcal{H}_i$, the history prior to time $i$. This allows us to calculate

$$\mathbf{E}\left(\Delta R_j^i \,\big|\, \mathcal{H}_i\right) = \sum_{\ell=1}^K p_\ell^i \left(c_\ell^i - c_j^i\right) = \left(\sum_{\ell=1}^K p_\ell^i c_\ell^i\right) - c_j^i$$

$$\mathbf{E}\left(\Delta A_j^i \,\big|\, \mathcal{H}_i\right) = p_j^i \left(1-b_j^i\right) \frac{c_j^i}{p_j^i} = \left(1-b_j^i\right) c_j^i$$

By Lemma 7.1, $\Delta \Phi_j^i \leq \nabla \Phi_j(\widehat{\mathbf{C}}^{i-1}) \cdot \mathbf{E}\left(\Delta \widehat{\mathbf{C}}^i \,\big|\, \mathcal{H}_i\right) + \eta K$. An easy calculation now shows

$$\nabla \Phi_j(\widehat{\mathbf{C}}^{i-1}) \cdot \mathbf{E}\left(\Delta \widehat{\mathbf{C}}^i \,\big|\, \mathcal{H}_i\right) = (\mathbf{e}_j - f(\widehat{\mathbf{C}}^{i-1})) \cdot \sum_{\ell=1}^K b_\ell^i c_\ell^i \mathbf{e}_\ell$$

$$= b_j^i c_j^i - \sum_{\ell=1}^K p_\ell^i b_\ell^i c_\ell^i.$$

Combining the above, we have

$$\mathbf{E}\left(\Delta R_j^i + \Delta \Phi_j^i + \Delta A_j^i \,\big|\, \mathcal{H}_i\right) \leq \sum_{\ell=1}^K (1-b_\ell^i) c_\ell^i p_\ell^i + \eta K.$$

Consequently,
$$\sum_{i=1}^{T} \mathbf{E}\left(\Delta R_j^i + {\Delta\Phi_j}^i + \Delta A_j^i \,\big|\, \mathcal{H}_i\right) \le \sum_{i=1}^{T}\sum_{\ell=1}^{K} (1-b_\ell^i) c_\ell^i p_\ell^i + \eta K T.$$

Since dropping the summands for which $p_\ell^i \le \eta$ can decrease the total by at most $\eta K T$, we can rewrite this as
$$\sum_{i=1}^{T} \mathbf{E}\left(\Delta R_j^i + {\Delta\Phi_j}^i + \Delta A_j^i \,\big|\, \mathcal{H}_i\right) \le \sum_{(i,\ell)\in S} c_\ell^i p_\ell^i + 2\eta K T, \qquad (3)$$
where the index set $S$ is defined by
$$S := \{(i,\ell) \colon b_\ell^i = 0 \text{ and } p_\ell^i > \eta.\}$$

Let $\alpha, \beta > 0$. Henceforth, our goal will be to prove an upper bound on
$$\mathbf{Pr}\left(\sum_{(i,\ell)\in S} c_\ell^i p_\ell^i \ge \alpha + \beta\right).$$

To this end, we will assume without loss of generality that the adversary is such that
$$\sum_{(i,\ell)\in S} c_\ell^i p_\ell^i \le \alpha + \beta,$$
in which case the probability we are trying to bound from above is
$$\mathbf{Pr}\left(\sum_{(i,\ell)\in S} c_\ell^i p_\ell^i = \alpha + \beta\right).$$

The intuition for why this assumption is valid is that, in those cases when the sum reaches $\alpha + \beta$, the adversary has already succeeded in his goal, so may as well stop.

For each $(i,\ell) \in S$, let $\chi_\ell^i$ denote the indicator random variable for the event that arm $\ell$ is chosen in round $i$. For $1 \le i \le T$, let
$$Z^i = \sum_{\ell:\, (i,\ell)\in S} c_\ell^i (p_\ell^i - \chi_\ell^i).$$

Note that $Z^i$ takes values in $[-1, 1]$ and that $Z^1, \ldots, Z^T$ is a martingale difference sequence. An easy calculation shows that
$$\mathbf{Var}\left(Z^i \,\big|\, \mathcal{H}_i\right) = \sum_{\ell:(i,\ell)\in S} (c_\ell^i)^2 p_\ell^i - \left(\sum_{\ell:(i,\ell)\in S} c_\ell^i p_\ell^i\right)^2 \le \sum_{\ell:(i,\ell)\in S} c_\ell^i p_\ell^i,$$
and hence that
$$\sum_{i=1}^{T} \mathbf{Var}\left(Z^i \,\big|\, Z^1, \ldots, Z^{i-1}\right) \le \sum_{(i,\ell)\in S} c_\ell^i p_\ell^i \le \alpha + \beta.$$

By McDiarmid's inequality, Theorem 5.1, applied to $Z = \sum_{i=1}^{T} Z^i$, with $a = \alpha$, $v = \alpha + \beta$ and $b = 1$, we have
$$\mathbf{Pr}(Z \geq \alpha) \leq \exp\left(-\frac{\alpha^2}{2\beta + 8\alpha/3}\right).$$

On the other hand,
$$\sum_{(i,\ell) \in S} c_\ell^i p_\ell^i = Z + \sum_{(i,\ell) \in S} p_\ell^i \frac{c_\ell^i \chi_\ell^i}{p_\ell^i}$$
$$= Z + \sum_{(i,\ell) \in S} p_\ell^i \Delta A_\ell^i$$
$$\leq Z + \sum_{(i,\ell) \in S} g(A_\ell^{i-1}) \Delta A_\ell^i \qquad \text{by definition of } S.$$

Hence,
$$\mathbf{Pr}\left(\sum_{(i,\ell) \in S} c_\ell^i p_\ell^i \geq \alpha + \beta\right) \leq \mathbf{Pr}(Z \geq \alpha) + \mathbf{Pr}\left(\sum_{(i,\ell) \in S} g(A_\ell^{i-1}) \Delta A_\ell^i \geq \beta\right)$$
$$\leq \exp\left(-\frac{\alpha^2}{2\beta + 8\alpha/3}\right) + \mathbf{Pr}\left(\sum_{(i,\ell) \in S} g(A_\ell^{i-1}) \Delta A_\ell^i \geq \beta\right). \qquad (4)$$

We claim that when $\beta = 3\theta$, that the last term equals zero. To see this, first note that the inequality $g(A_\ell^i) \geq \frac{2}{3} g(A_\ell^{i-1})$ always holds. Since, for each $1 \leq \ell \leq K$, $g(A_\ell^i)$ is always a decreasing sequence in $i$, it follows that

$$\sum_{(i,\ell) \in S g(A_\ell^{i-1}) > \eta} g(A_\ell^{i-1}) \Delta A_\ell^i \leq \frac{3}{2} \sum_{(i,\ell) \in S g(A_\ell^{i-1}) > \eta} g(A_\ell^i) \Delta A_\ell^i$$
$$\leq \frac{3}{2} \sum_{\ell=1}^{K} \sum_{i=1}^{T} \frac{1}{K(1 + A_\ell^i/\theta)^{3/2}} \Delta A_\ell^i$$
$$\leq \frac{3K}{2} \int_0^{A_{\text{crit}}} \frac{1}{K(1 + x/\theta)^{3/2}} dx \qquad \text{(since the integrand is a decreasing function)}$$
$$\leq \frac{3}{2} \int_0^{\infty} \frac{1}{(1 + x/\theta)^{3/2}} dx$$
$$= 3\theta \frac{-1}{(1 + x/\theta)^{1/2}} \bigg|_0^{\infty}$$
$$= 3\theta.$$

This establishes our claim. Setting $\alpha = \theta$, and combining this with (3) and (4) allows us to conclude

$$\mathbf{Pr}\left(\sum_{i=1}^{T} \mathbf{E}\left(\Delta R_j^i + \Delta \Phi_j^i + \Delta A_j^i \,\big|\, \mathcal{H}_i\right) > 6\theta\right) \leq \exp(-3\theta/26).$$

$\square$

# 8 An adaptive cost schedule: Proof of Theorem 1.2

In this section, we present a simple family of adaptive cost schedules. We also outline the proof of Theorem 1.2, showing that no matter how the parameters of the **Exp3** algorithm of Auer *et al.* [1] are set, the expected regret will be $\Omega(T^{2/3})$ for some cost schedule from this family.

Let $K = 2$. (Adding extra arms which always have cost 1 can only increase the regret.) Let $\mathcal{A}$ be any algorithm for the gambler. For every $\alpha \in [0, 1]$, let $\mathcal{V}(\mathcal{A}, \alpha)$ be the following adaptive strategy for setting costs. Let $p$ denote the conditional probability that $\mathcal{A}$ chooses arm 1 at time $t$, given the history of the game on steps $1, \ldots, t-1$. Then $\mathcal{V}(\mathcal{A}, \alpha)$ sets the cost vector $\mathbf{c}^t$ as follows:

$$\mathbf{c}^t := \begin{cases} \mathbf{e}_2 & \text{if } p < \alpha \\ \mathbf{e}_1 & \text{otherwise.} \end{cases}$$

The motivation behind this adaptive method of setting costs is to encourage algorithm $\mathcal{A}$ to always move its probability distribution towards (and perhaps beyond) $\alpha$. Note that this works in the case of all the multiplicative weights-based algorithms discussed in this paper; whenever these algorithms see a cost of 0, they keep the weight for that arm fixed, but when they see a cost of 1, they decrease that arm's weight.

**Observation 8.1.** *When **Exp3** is run against adaptive costs from $\mathcal{V}(\mathbf{Exp3}(\gamma, \eta), \alpha)$ for infinitely many steps, the sequence of probabilities $p^t$ is uniquely determined modulo consecutive repetitions of a single value. Whenever $p^t < \alpha$, $p^{t+1} \geq p^t$, and whenever $p^t \geq \alpha$, $p^{t+1} \leq p^t$.*

*Proof.* At each time step, there are only two random possibilities: a 0 is observed, or a 1. When a 0 is observed, the algorithm, and hence the cost sequence, behaves exactly the same as if the round had not occurred. This results in a repetition. When a 1 is observed, the algorithm shifts $p$ towards (and perhaps beyond) $\alpha$, by an amount which is uniquely determined. □

We now argue that, for every setting of the parameters $\gamma, \eta$, of the **Exp3** algorithm of Auer *et al.* [1, 2], there exists an $\alpha \in [0, 1]$ such that against the adaptive cost schedule $\mathcal{V}(\mathbf{Exp3}(\gamma, \eta), \alpha)$, the expected regret of $\mathbf{Exp3}(\gamma, \eta)$ is $\Omega(T^{2/3})$. (Note that $\gamma, \eta$ may depend on $T$.)

*Proof sketch for Theorem 1.2.* Suppose **Exp3** is run against adaptive costs from $\mathcal{V}(\mathbf{Exp3}(\gamma, \eta), \alpha)$. Then, clearly,

- the loss of the algorithm equals the number of steps when a 1 is observed.
- The loss of arm 1 equals the number of steps when $p < \alpha$,
- and the loss of arm 2 equals the number of steps when $p \geq \alpha$.

We will focus on the case when $\gamma = \eta = \Theta(T^{-1/2})$, under which parameters **Exp3** has expected regret $O(\sqrt{T})$ against any non-adaptive adversary. Let $\alpha = 3\gamma$. In this case, it is not hard to see that, with high probability, $p$ will lie in the interval $[2\gamma, 4\gamma]$ for almost all $T$ rounds of the game.

Moreover, $p$ will cross from greater than $\alpha$ to less than $\alpha$ approximately $\alpha T/2$ times, taking one big step down, and approximately $1/\alpha$ small steps up (the exact number is determined, but will not concern us). Let us look in more detail at what happens during each such "loop traversal."

After each downward crossing of $\alpha$, the algorithm has probability about $1 - \alpha$ to see a 1 at each time step. Since about $1/\alpha$ upward moves must be made to cross $\alpha$ again, this implies that the number of steps before the next upward crossing of $\alpha$ has expectation about $1/\alpha$ and variance $O(1)$. After each upward crossing of $\alpha$, the algorithm has probability approximately $\alpha$ to see a 1 at each time step. Since only one 1 downward move is needed to cross $\alpha$ again, this implies that the number of steps before the next downward crossing of $\alpha$ has expectation about $1/\alpha$ and variance $\Theta(1/\alpha^2)$.

These facts together imply that, in any single loop, the change in arm-specific regret $\Delta R_1$ is a random variable with mean $O(1)$ and variance $O(1)$. On the other hand, $\Delta R_2$ is a random variable with mean $O(1)$ and variance $\Theta(1/\alpha^2)$ (a shifted exponential distribution).

If instead of being played for a fixed number of time steps, the game were played for exactly $\alpha T/2$ complete loops (and ignoring the contribution of the few steps before the first downward crossing of $\alpha$), then the total arm-specific regrets would be the sum of $\alpha T/2$ independent trials of these two respective random variables. In this case, the expectations $\mathbf{E}(R_1)$ and $\mathbf{E}(R_2)$ would each be $\Theta(\alpha T)$, however the variances would be $\Theta(\alpha T)$ and $\Theta(T/\alpha)$, respectively.

More to the point, $|R_1|$ is almost always $O(\alpha T)$, but $R_2$ is $\Omega(\sqrt{T/\alpha})$ with constant probability. To see the latter, note that the loss of arm 2 is less than or equal to $L$ if and only if, in $L$ coin flips with probability $\alpha$ of heads, at least $\alpha T/2$ heads come up. The probability of this equals

$$\sum_{j \geq \alpha T/2} \binom{L}{j} \alpha^j (1-\alpha)^{L-j}.$$

The desired bound can, with a little work, be inferred using Stirling's formula, or from basic properties of binomial distributions; we omit the details from this sketch.

Now, since $R_1$ is almost always greater than $-C\alpha T \approx -C\sqrt{T}$ and $R_2$ is with constant probability greater than $\sqrt{T/\alpha} \approx T^{3/4}$, it follows that $\mathbf{E}(R) = \Omega(\sqrt{T/\alpha}) = \Omega(T^{3/4})$.

Analogous arguments can be made for all other values of $\gamma, \eta$, proving an $\Omega(T^{2/3})$ lower bound in general. Note that if $\gamma > T^{-1/3}$, one should set $\alpha < \gamma$, in which case the argument is different but easier, and the lower bound is $\gamma T$. □

## 9 Applications

The $k$-armed bandit has been used as a model for a wide variety of online decision-making problems, such as combining expert advice, portfolio balancing, machine learning (boosting), network routing, and sequential auctions (among others). In many of these contexts, it would be desirable to provide a high-probability guarantee on the actual regret, rather than the expected regret, and/or to relax the assumption that the decisions made by the algorithm have no effect on the distribution of subsequent incurred costs. The "Accounts" algorithm provides both of these features.

We plan to add a more detailed description of some of these applications in a later version of this paper.

## Acknowledgements

We would like to thank Avrim Blum, Adam Kalai, Bobby Kleinberg and Alistair Sinclair for proofreading preliminary drafts of this paper and for helpful discussions.